\documentclass[aps,prb,reprint,showpacs,groupedaddress]{revtex4-1}
\usepackage{graphicx}
\usepackage{dcolumn}
\usepackage{bm}

\begin{document}


\title{Transmission enhancement in three-dimensional rolled-up plasmonic metamaterials containing optically active quantum wells}

\author{Andreas Rottler} \email{arottler@physnet.uni-hamburg.de}
\affiliation{Institut f\"ur Angewandte Physik und
Mikrostrukturforschungszentrum, Universit\"at Hamburg,
Jungiusstrasse 11, D-20355 Hamburg, Germany}
\author{Stephan Schwaiger}
\affiliation{Institut f\"ur Angewandte Physik und
Mikrostrukturforschungszentrum, Universit\"at Hamburg,
Jungiusstrasse 11, D-20355 Hamburg, Germany}
\author{Aune Koitm\"ae}
\affiliation{Institut f\"ur Angewandte Physik und
Mikrostrukturforschungszentrum, Universit\"at Hamburg,
Jungiusstrasse 11, D-20355 Hamburg, Germany}
\author{Detlef Heitmann}
\affiliation{Institut f\"ur Angewandte Physik und
Mikrostrukturforschungszentrum, Universit\"at Hamburg,
Jungiusstrasse 11, D-20355 Hamburg, Germany}
\author{Stefan Mendach}
\affiliation{Institut f\"ur Angewandte Physik und
Mikrostrukturforschungszentrum, Universit\"at Hamburg,
Jungiusstrasse 11, D-20355 Hamburg, Germany}

\date{\today}

\begin{abstract}
We investigate three-dimensional rolled-up metamaterials containing optically active quantum wells and metal gratings supporting surface plasmon polarition resonances. Finite-difference time-domain simulations show that by matching the surface plasmon polarition resonance with the active wavelength regime of the quantum well a strong transmission enhancement is observed when illuminating the sample with p-polarized radiation. This transmission enhancement is further increased by taking advantage of the Fabry-Perot resonances of the structure.
\end{abstract}


\maketitle
\section{Introduction}

The emerging field of metamaterials and the realization of materials exhibiting a negative index of refraction (NIM) in the past ten years opened the way for various fascinating applications. With these novel materials physical applications like cloaking devices \cite{Schurig2006, Cai2007, Ergin2010} and optical imaging beyond the diffraction limit \cite{Liu2007, Zhang2008, Kawata2009} have become possible. A metamaterial consists of artificial structures which are much smaller than the wavelength of the operating incident radiation. NIMs have been realized by metal-dielectric structures e.g. split-ring resonators deposited on dielectrics \cite{Shelby2001, Katsarakis2005} or fishnet structures \cite{Zhang2005b, Dolling2006, Xiao2009}. However, due to the electron-beam lithography fabrication process of these structures, the realization of three-dimensional metamaterials requires a sequentially stacking of single layers \cite{Dolling2007b, Liu2007a, Valentine2008}. Three-dimensional metamaterials consisting of multistacks of metal-dielectric layers can also be obtained by rolling-up strained layers into microrolls \cite{Prinz2000, Schumacher2005}. That opens up the application of such microrolls for example as hyperlenses \cite{Schwaiger2009}.\\
\indent The efficiency of metamaterials suffers from the absorption of radiation in the metal, caused by the finite imaginary part of the metals dielectric function. This causes strong energy dissipation and a rapid field decay. A possible solution to overcome this problem would be the incorporation of gain materials. Different approaches in the recent literature include dyes \cite{Xiao2010}, quantum dots \cite{Plum2009} and optically-pumped quantum wells that have been brought in close proximity to split-ring resonators \cite{Meinzer2010}. Semiconductor quantum wells show no bleaching and electrical pumping is well established for these structures. Recently it was demonstrated that the self-rolling process offers the possibility to easily include active quantum-wells in the semiconductor layer of three-dimensional metamaterials. It was found that such a novel structure shows enhanced light transmission upon optical pumping \cite{Schwaiger2011}. In the next step the concept of rolled-up nanotech allows the realization of three-dimensional metamaterials with sophisticated metallic nanostructures instead of planar metal layers which enables us to profit from plasmonic resonances.\\
\begin{figure}
\includegraphics[width=1.0\columnwidth]{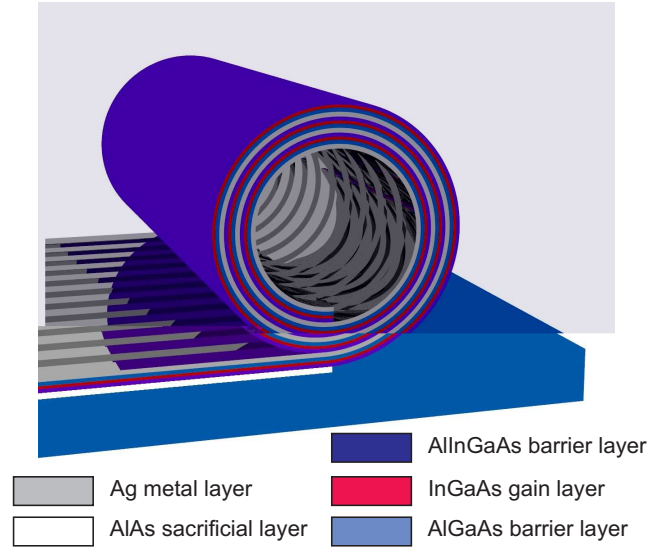}
\caption{(Color online) Sketch of a micoroll which can be fabricated by rolling-up strained layers. The tube wall represents a three-dimensional metamaterial consisting of a metal-semiconductor superlattice containing quantum wells and metal gratings.}
\end{figure}
In this letter we investigate alternating layers of amplifying quantum wells and metal gratings supporting surface plasmon polaritions (SPPs). Such structures can be realized in the walls of rolled-up metamaterials as sketched in Fig.~1. Experimentally one can prepare different type of layered systems rolling along or perpendicular to the grating stripes. Here we are interested in the fundamental effect of grating coupler enhanced gain and concentrate on a microroll with rolling direction along the grating stripes where we can precisely align the stripes on top of each other. Whereas in the case of a perpendicular rolling direction commensurability effects 
between the microroll perimeter and the grating period can occur. We present finite-difference time-domain (FDTD) simulations which show a transmission enhancement in the structures with metallic gratings much larger than observed for samples with flat silver surfaces.

\section{Structures and Simulation Method}

In Fig.~1 a sketch of a microroll is shown. The preparation of this structures is based on strain relaxation of pseudomorphically grown (AlIn)GaAs heterostructures, which roll up into a tube when released from the substrate by etching away an AlAs sacrificial layer (for details see \cite{Schwaiger2009}). The important ingredients for our investigation here is that in addition (i) an InGaAs gain layer and (ii) a grating is etched into the semiconductor layer and covered with Ag before the rolling-up process. As a result one obtains alternating layers of quantum wells and Ag gratings.\\
\begin{figure}
\includegraphics[width=1.0\columnwidth]{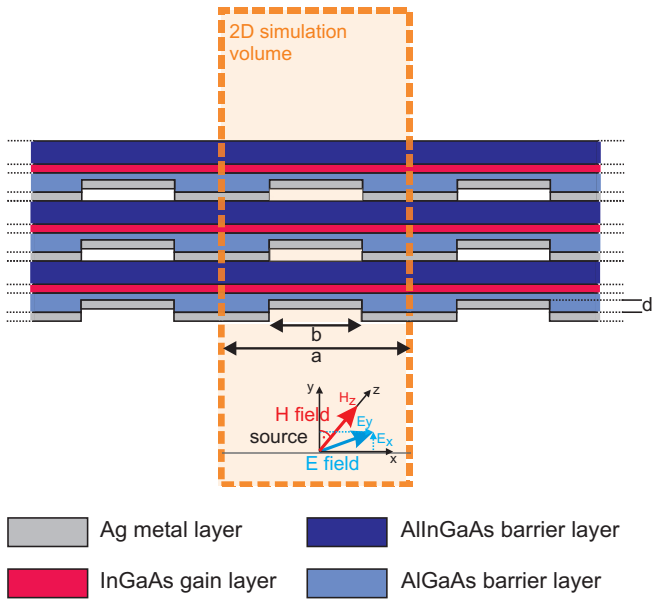}
\caption{(Color online) Scheme of the 2D simulation volume (orange box) with p-polarized fields. $a$ is the lattice constant, $b$ denotes the width of the etched region and $d$ is the grating depth.}
\end{figure}
We performed finite-difference time-domain simulations on the metamaterials by using the commercial software Lumerical FDTD Solutions \cite{Lumerical}. For the simulations we neglected the curvature of the actual structure and investigate flat structures in two dimensions. We simulated a grating unit cell (lattice constant $a$ and filling factor $t=b/a$ where $b$ denotes the width of the etched region) with three layers where we used periodic boundary conditions in x direction and perfectly matched layer boundary conditions in y direction (Fig.~2). We irradiated the investigated structures with a plane wave with zero incident angle. To fulfill the criterion for SPP excitation we assume that this plane wave has the H field $\vec{H}=(0,0,H_z)$ and the E field $\vec{E}=(E_x,0,0)$. Due to the periodic corrugation also y components of the electric field are induced. In the frequency spectrum the plane wave pulse has a Gaussian-shaped distribution with a maximum at $406\;$THz and a full-width-half-maximum (FWHM) of $385\;$THz. This corresponds to broadband excitation with a center wavelength of $\lambda=950\;$nm and a FWHM of $\Delta\lambda=900\;$nm. Inside the simulation volume we used an appropriate non-uniform mesh which has been optimized by convergence testing. The transmission spectrum was recorded in a distance of $d_{\mathrm{trans}}=0.6\;\mu$m in y direction behind the structure. For the dielectric functions we take polynomial fits on experimental data from Palik \cite{Palik1985}. The dielectric functions of the AlGaAs, the InGaAs and the AlInGaAs used in the simulations are based on the dielectric function of GaAs and were adjusted as follows: The percentage of In and Al in the semiconductor layer decreases and increases the bandgap of the composite semiconductor which we assume to be Al$_{26}$Ga$_{74}$As, In$_{16}$Ga$_{84}$As and Al$_{20}$In$_{16}$Ga$_{64}$As. In first approximation this shift of the bandgap is linear causing the dielectric function to change slightly. This effect was taken into account by dislocating the dielectric function of GaAs linearly,
$$\epsilon_{\mathrm{SC}}(\hbar \omega) = \epsilon_{\mathrm{GaAs}}\left(\hbar \omega- \Delta E_{\mathrm{In}53} \frac{P_{\mathrm{In}}}{0.53} + \Delta E_{\mathrm{Al}30} \frac{P_{\mathrm{Al}}}{0.3}\right)$$
where $P_{\mathrm{In}}$ and $P_{\mathrm{Al}}$ are the percentages of In and Al in the semiconductor, respectively. The difference of the bandgap of GaAs and In$_{53}$Ga$_{47}$As \cite{Goetz1983} is $\Delta E_{\mathrm{In}53}=0.7\;$eV and the difference of the bandgap of Al$_{30}$Ga$_{70}$As and GaAs \cite{Monemar1976} is $\Delta E_{\mathrm{Al}30}=0.45\;$eV.\\
\indent The amplification effect of the InGaAs quantum well can be described by a negative imaginary part of the refractive index. In the simulations the dielectric function $\epsilon_{\mathrm{gain}}$ of this gain layer was modelled qualitatively as a Lorentz oscillator according to \cite{Govyadinov2007}:
$$\epsilon_{\mathrm{gain}}=\epsilon_{\mathrm{InGaAs}}-\frac{\omega_0^2\xi}{\omega_0^2-i\omega\gamma-\omega^2},$$
with $\xi$ as the Lorentz oscillator strength, $\gamma$ the damping frequency and $\omega_0$ as the resonant frequency of the quantum well.

\section{Parameter calibration}

The beneficial exploitation of the SPP excitations on the transmission requires that the lattice constant of the metallic grating is chosen appropriately to adjust the energy of the SPP resonance to the active wavelength regime of the of the quantum well. Furthermore it is favorable to tune the SPP resonance energy to the Fabry-Perot transmission peak of the system arising from the layered semiconductor/metal system with its high index of refraction.
\begin{figure*}
\includegraphics{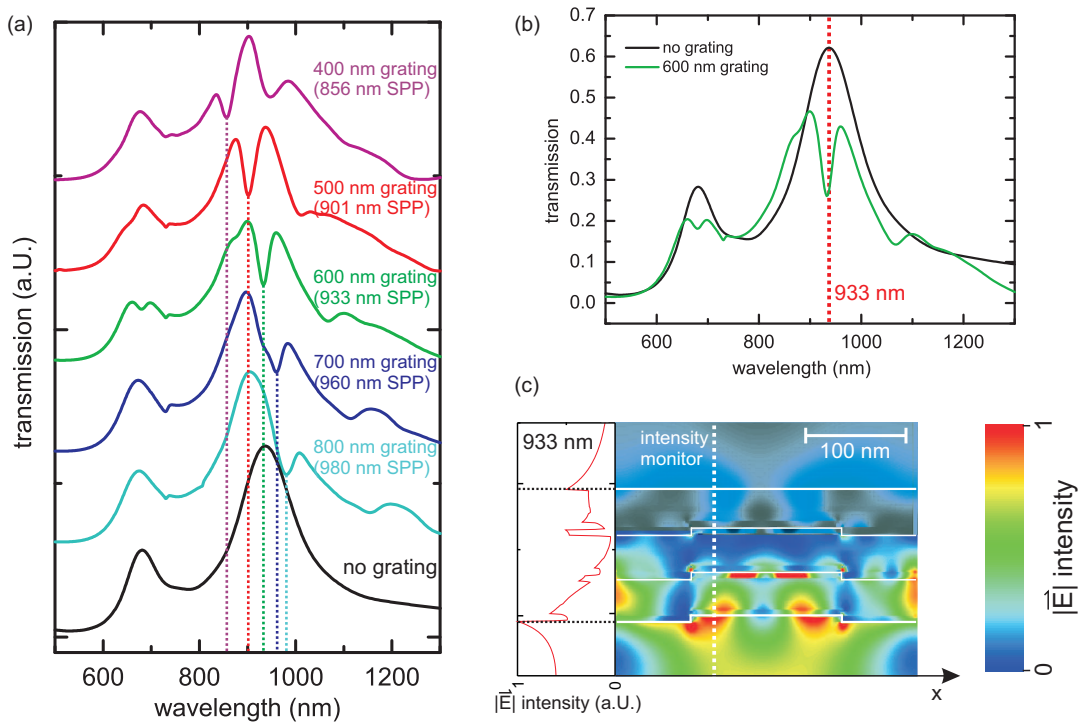}
\caption{(Color online) (a) Transmission spectrum of a three-layer structure of $10\;$nm Ag, $26\;$nm AlInGaAs, $9\;$nm InGaAs and $25\;$nm AlGaAs with different metal grating periods and $10\;$nm lattice depth in comparison to flat layers (black line). (b) Transmission spectrum of an $a=600\;$nm grating (green line) together with the transmission spectrum of flat layers (black line). The quantum well wavelength is marked with a red dashed line. (c) shows the two-dimensional electric-field distribution in the regime of the $\lambda=933\;$nm SPP mode. Adjacent we plot the one-dimensional intensity profile originating from the white dashed line (the intensity axis have the same linear scaling in both cases).}
\end{figure*}
The optically active wavelength regime of the quantum well can in fact be tailored by the amount of Indium used during sample preparation. In the simulations presented in this letter we consider a quantum well emitting at $\lambda=933\;$nm. The Fabry-Perot resonance peaks can be shifted by varying the individual layer thicknesses. The black line in Fig.~3(a) shows the simulated transmission spectrum for a three-layer structure with each layer consisting of $10\;$nm Ag, $26\;$nm AlInGaAs, $9\;$nm InGaAs and $25\;$nm AlGaAs. We see that, for this dimensions, the structure exhibits a pronounced Fabry-Perot peak at the desired wavelength of $\lambda=933\;$nm. The appropriate grating was examined by performing simulations with various lattice constants. In Fig.~3(a) we display the transmission spectra for different gratings in comparison with the spectrum of unstructured layers. We first choose a grating filling factor $t=0.5$, a depth $d=10\;$nm and an Ag thickness $t_{\mathrm{Ag}}=10\;$nm. One can see that for an $a=600\;$nm grating a deep SPP resonance is located at the desired wavelength of $\lambda=933\;$nm. For smaller grating periods the SPP resonance shifts to shorter wavelengths ($\lambda=856\;$nm for an $a=400\;$nm grating and $\lambda=901\;$nm for an $a=500\;$nm grating) whereas for larger grating periods the SPP-resonance wavelength is longer ($\lambda=960\;$nm for an $a=700\;$nm grating and $\lambda=980\;$nm for an $a=800\;$nm grating). Variations in the lattice depth do not affect the position of the SPP resonance but a smaller lattice depth causes less pronounced resonances. Moreover, several other SPP excitations can be observed, however, they are considerably less pronounced. From the observed resonance positions we can reconstruct a dispersion relation $\omega_{\mathrm{SP}}=\omega_{\mathrm{SP}}(k_x=n\cdot 2\pi /a)$ where $n$ is the diffraction order. We find that the pronounced minima, starting for $a=800\;$nm at $\lambda=980\;$nm, follow a dispersion $\omega_{\mathrm{SP,}a}$ with $n=1$. The minima in the regime between $\lambda=650\;$ to $\lambda=750\;$nm follow the same dispersion with $n=2$, thus they represent a higher diffraction order. This dispersion $\omega_{\mathrm{SP,}a}$ lies inbetween the SPP dispersion of a flat-Ag/vacuum interface and a flat-Ag/GaAs interface. This is what we actually expect for our complex semiconductor/metal structures including vacuum gaps. We further expect even different SPP branches due to the asymmetric arrangement and coupling between the three metal layers. Indeed, the resonances starting for $a=800\;$nm at $\lambda=1155\;$nm follow a dispersion $\omega_{\mathrm{SP,}b}$ that lies energetically below $\omega_{\mathrm{SP},a}$. Our studies showed that the variation of the lattice filling factor (we simulated filling factors between $t=0.2$ and $t=0.8$) does effect the position of the SPP resonance only slightly by about $\Delta\lambda=20\;$nm. In Fig.~3(b) we independently show the transmission spectra for the above structure without and with an $a=600\;$nm grating. It can be seen clearly that the SPP resonance and the Fabry-Perot resonance peak are matched to the active wavelength regime of the quantum well ($\lambda=933\;$nm). In Fig.~3(c) we plot the two-dimensional electric-field distribution of the $\lambda=933\;$nm SPP mode (adjacent we show the one-dimensional intensity profile corresponding to the white dashed line). We recognize that the intensity profile exhibits a large intensity inside the structure in particular at the quantum well locations. We note that in two-dimensional metamaterials the gain material has to be brought in very close proximity to the metallic structure (see, e.g., reference \cite{Meinzer2010}) which can cause manufacturing problems. In our three-dimensional case the mode pattern allows us to use thick barrier layers which are necessary to maintain the optical quality of the quantum well and still have a strong field-coupling to the quantum well. 
\begin{figure}
\includegraphics[width=0.9\columnwidth]{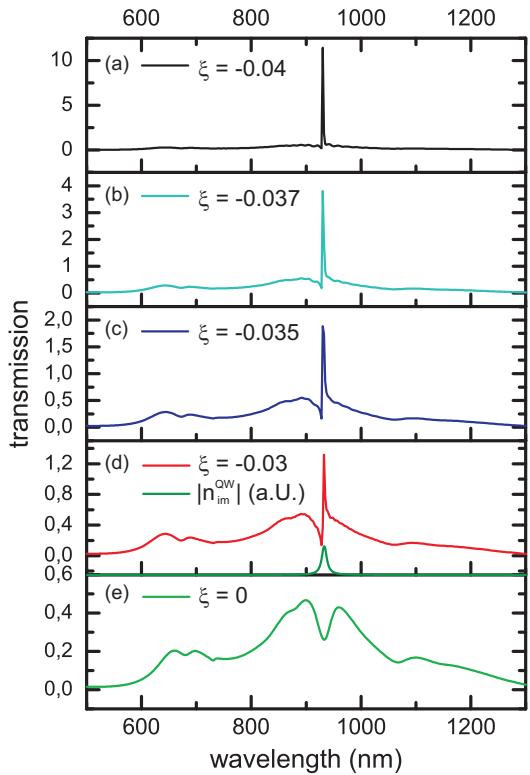}
\caption{(Color online) (a)-(d) Transmission versus wavelength spectra for different values of gain and a grating period of $a=600\;$nm for illumination with a p-polarized wave. The InGaAs gain layer has a thickness of $9\;$nm. (d) also shows the imaginary part of the refractive index of the gain layer. (e) Transmission spectrum without gain (compare with Fig.~3(b)) for illumination with a p-polarized wave.}
\end{figure}
\begin{figure}
\includegraphics[width=0.9\columnwidth]{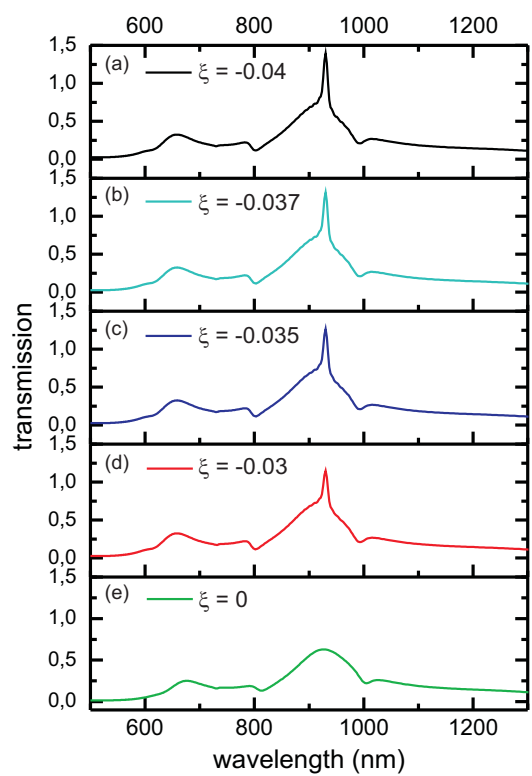}
\caption{(Color online) (a)-(d) Transmission versus wavelength spectra for different values of gain and a grating period of $a=600\;$nm for illumination with a s-polarized wave. The quantum well has a thickness of $9\;$nm. (e) Transmission spectrum without gain for illumination with a s-polarized wave. A pronounced SPP resonance at $\lambda=933\;$nm wavelength is not observed (compare with Fig.~4(e)).}
\end{figure}

\section{Results and Discussion}

The Lorentz oscillator strength in the gain layer was adjusted by investigating the transmission enhancement on a single layer: In previous experiments without grating it was shown, that for single layers with and without a gain layer, which exhibit transmissions $T$ and $T+\Delta T$, respectively, a transmission enhancement of about $\Delta T/T=2\%$ is easily realizable \cite{Schwaiger2011}. We performed simulations where we adjusted the Lorentz-oscillator strength $\xi$ to $\xi=-0.03, -0.035, -0.037, -0.04$. This corresponds to single layer transmission enhancements of $\Delta T/T=2\%, 3\%, 4\%, 5\%$ and imaginary parts of the Lorentz-oscillators refractive index of $n_{\mathrm{im}}^{\mathrm{QW}}=-0.37, -0.43, -0.45$ and $-0.50$ at $\lambda=933\;$nm, respectively. The results are plotted in Fig.~4. Figure~4(e) shows the transmission spectrum with no gain as a reference. Figures~4(a)-(d) depicts the transmission versus wavelength spectra for the above mentioned values of $\xi$. It is observed that the transmission at the SPP resonance is drastically enhanced with increasing gain. For $\xi=-0.03$ (corresponding to $n_{\mathrm{im}}^{\mathrm{QW}}=-0.37$ and $\Delta T/T=2\%$ for an unstructured single layer film) we already obtain a transmission of more than $1$ at the wavelength of $933\;$nm. With increasing gain the transmission at this wavelength increases up to 10 for $\xi=-0.04$ ($n_{\mathrm{im}}^{\mathrm{QW}}=-0.50$, $\Delta T/T=5\%$). Figure~3(c) shows that there is a high intensity in all three layers. This indicates that the multilayered systems that we can realize with the microroll technique is indeed quite helpful. We have actually performed simulations, where we selectively turn on the gain in different layers and it turned out that the effects on the transmission are considerably smaller. It is therefore advantageous to use a structure with three or even more layers. We like to note that, as observed in Fig.~4, the transmission enhancement has actually a Fano-type shape with a minimum at the low wavelength side. This indicates that the negative imaginary part of the gain material shifts the SPP resonance position, i.e. there is an increasing interaction between the SPP and the gain layer.\\
While the excitation of SPPs is linked to illumination with p-polarized light we also performed for comparison simulations in s polarization. The transmission versus wavelength spectra of this run are presented in Fig.~5. Figures~5(a)-(d) show the spectra for Lorentz-oscillator strengths $\xi=-0.03, -0.035, -0.037, -0.04$ corresponding to $\Delta T/T=2\%, 3\%, 4\%$ and $5\%$, respectively. Figure~5(e) depicts the transmission spectrum without gain. As expected, there are no SPP resonances. Turning on the gain leads to an increasing transmission at the resonant frequency of the quantum well. However, in the s-polarized case a strong transmission enhancement as in Fig.~4(a)-(d) is not observed. Our simulations show that the increased transmission for s polarization is comparable to simulations with flat layers instead of a grating. This observation supports that the SPP resonances are crucial for the extraordinary transmission. Further simulations show that the strength of the transmission enhancement as seen in Fig.~4 is strongly dependent on the interplay of the Fabry-Perot resonance, the SPP resonance and the active wavelength regime of the quantum well.

\section{Conclusions}

In conclusion we demonstrated that SPP resonances on metallic gratings embedded into three-dimensional metamaterials containing quantum structures can lead to a strong transmission enhancement. The dimensions of the structures and the grating period have to be chosen such that the wavelength of the SPP resonance spectrally and spatially matches the emission of the quantum well. Moreover our simulations show that it is desirable to also shift the Fabry-Perot resonance peak to the SPP resonance.\\

\begin{acknowledgments}
The authors thank Markus Br\"oll and Jens Ehlermann for fruitful discussions and gratefully acknowledge financial support of the Deutsche Forschungsgemeinschaft via the Graduiertenkolleg 1286 ``Functional Metal-Semiconductor Hybrid Systems''.
\end{acknowledgments}

\end{document}